\documentclass[a4paper,twocolumn,showpacs,nofootinbib,floatfix,superscriptaddress,prc]{revtex4}
\usepackage{graphicx}
\usepackage{amsfonts}
\usepackage{amsmath}
\usepackage{amsbsy}
\usepackage{longtable}
\usepackage{bm}
\usepackage{hyperref}
\usepackage{float}

\def\be{\begin{equation}}
\def\ee{\end{equation}}
\def\bea{\begin{eqnarray}}
\def\eea{\end{eqnarray}}

\begin{document}

\title{Extension of relativistic dissipative hydrodynamics to third order}

\author{Andrej\ El\footnote{el@th.physik.uni-frankfurt.de}}
\affiliation{Institut f\"ur Theoretische Physik, 
 Goethe-Universit\"at, 
Max-von-Laue-Str.\ 1, D-60438 Frankfurt am Main, Germany}

\author{Zhe\ Xu}
\affiliation{Institut f\"ur Theoretische Physik, 
 Goethe-Universit\"at, 
Max-von-Laue-Str.\ 1, D-60438 Frankfurt am Main, Germany}

\author{Carsten\ Greiner}
\affiliation{Institut f\"ur Theoretische Physik, 
 Goethe-Universit\"at, 
Max-von-Laue-Str.\ 1, D-60438 Frankfurt am Main, Germany}

\begin{abstract}
Following the procedure introduced by Israel and Stewart, we expand the entropy current up to the
third order in the shear stress tensor $\pi^{\alpha\beta}$ and derive a novel third-order
evolution equation for $\pi^{\alpha\beta}$. This equation is solved for the one-dimensional 
Bjorken boost-invariant expansion. The scaling solutions for various values of the shear viscosity
to the entropy density ratio $\eta/s$ are shown to be in very good agreement with those obtained
from kinetic transport calculations. For the pressure isotropy starting with 1 at $\tau_0=0.4
fm/c$, the third-order corrections to
Israel-Stewart theory are approximately 10\% for $\eta/s=0.2$ and more than a factor of 2 for 
$\eta/s=3$. We also estimate all higher-order corrections to
Israel-Stewart theory and demonstrate their importance in describing highly viscous matters.

\end{abstract}

\date{\today}

\maketitle
A causal theory of relativistic dissipative hydrodynamics was first 
formulated by Israel and Stewart \cite{IS} and has been successfully 
applied to study and understand a wide range of phenomena observed 
in ultrarelativistic heavy-ion 
collisions \cite{Heinz:2009xj,Romatschke:2009im,Teaney:2009qa}.
The success may lie on the speculation that the viscosity of the hot dense
matter created experimentally is really small. Since the Israel-Stewart 
theory is a second-order theory, which neglects the higher orders
in viscous stress, it is natural to look for its limit of applicability. 
Especially for one-dimensional expansion with the Bjorken
boost invariance \cite{Bjorken:1982qr} the Israel-Stewart theory has led to
a reheating of the 
expanding medium \cite{M04} and to a negative pressure \cite{Mart09}, 
if the viscosity is large or equivalently the starting time of expansion
is small due to the scaling behavior of the solutions.
These effects are unphysical and thus show the breakdown of the 
Israel-Stewart theory. 
Moreover, in Refs. \cite{HM09,El09}, the scaling solutions of the
Israel-Stewart hydrodynamic equations are compared with those obtained
from kinetic transport calculations. The breakdown of the second-order 
theory has been found at $\eta/s > 0.5$. From comparisons of dissipative 
hydrodynamics calculations to the elliptic flow measurements one infers that 
$\eta/s\sim 0.5$ is an upper limit for the shear viscosity to entropy density ratio of a
quark-gluon plasma 
(QGP) \cite{GA06,D07,Rom08}. It is thus of interest to extend the second-order Israel-Stewart theory
in order to include higher-order corrections and then to quantify their effects on observables.

Using the Grad's 14-moment method and neglecting bulk pressure and heat flow, the off-equilibrium
distribution function is given by\cite{El09,M07}
\begin{equation}
f(x,p)= f_0(x,p) (1+\phi) \approx 
f_0(x,p)\left(1+C_0\pi_{\mu\nu}p^\mu p^\nu \right).\
\label{Grad}
\end{equation}
$f_0(x,p)$ is the Boltzmann distribution in kinetic equilibrium
\begin{equation}
f_0(x,p)=(2\pi)^{-3}\lambda exp\left(-\frac{p_\mu u^\mu}{T}\right) ,\
\label{feq}
\end{equation}
where $\lambda$ denotes the fugacity describing the chemical equilibration, $T$ is the local
temperature and $u^\mu$ is flow velocity. The expression in Eq.(\ref{Grad}) is first order in
$\pi^{\mu\nu}=T^{\mu\nu}-T^{\mu\nu}_{eq}$
 which is the deviation of the energy-momentum tensor from its equilibrium form. We consider a
system of massless particles which leads to a vanishing bulk pressure. Using the Landau
matching conditions $u_\mu u_\nu T^{\mu\nu} = u_\mu N^\mu = 0$  we obtain $C_0=3/(8eT^2)$
\cite{El09}. The Landau matching conditions
are equivalent to $e=e_{eq}$ and $n=n_{eq}$ in the comoving frame, where $e_{eq}=3\lambda
T^4/\pi^2$ and
$n_{eq}=\lambda T^3/\pi^2$ denote the equilibrium values for energy and particle densities according
to Eq.(\ref{feq}). This allows us to define the effective temperature
$T=e/(3n)$. 

We follow the approach introduced by Israel and Stewart and use the entropy principle to derive an
evolution equation for the shear tensor. The entropy current $s^\mu$ can be calculated according to
the kinetic definition:
\begin{equation}
s^\mu = -\int \frac{d^3 p}{E}  p^\mu f (\ln f - 1) \,. 
\end{equation}
$\ln (f)$ will be expanded to the third order in $\phi\approx C_0\pi_{\mu\nu}p^\mu p^\nu$ [see
Eq.(\ref{Grad})]. We obtain 
\begin{eqnarray}
s^\mu &\approx& -\int \frac{d^3 p}{E} \ f_0 p^\mu \left( \ln f_0 - 1 + \phi + \phi \ln f_0 +
\frac{\phi^2}{2}-\frac{\phi^3}{6} \right) \nonumber \\
&=& s_0 u^\mu - \frac{\beta_2}{2T} \pi_{\alpha\beta}\pi^{\alpha\beta} u^\mu -
\frac{8}{9}\frac{\beta_2^2}{T}
\pi_{\alpha\beta}\pi^\alpha_\sigma \pi^{\beta\sigma} u^\mu ,\
\label{entropy_3rd_order}
\end{eqnarray}
where $s_0=-\int d^3p \ f_0 ( \ln f_0 - 1)=4n-n \ln \lambda$ is the
entropy density in kinetic equilibrium and $\beta_2=9/(4e)$.  The value of
$\beta_2$ we obtained is just the same as derived in the Israel-Stewart theory \cite{IS,M04}.

Up to second order in $\pi^{\alpha\beta}$ the entropy current in Eq.(\ref{entropy_3rd_order}) is the
same as in Israel-Stewart's approach\cite{IS,M04,A07}. However, it looks different from the
expression found in Refs. \cite{Bh08,Rom10} for conformal fluids. There, the entropy current is
constructed to include all possible terms up to the second order in gradient of the flow velocity.
In general,
the shear tensor $\pi^{\alpha\beta}$ can be expanded in terms of gradient of the flow velocity by
introducing transport coefficients. Up to the second order the expansion is explicitly given in
Ref. \cite{Rom09} for conformal fluids. If we insert this expansion into
Eq.(\ref{entropy_3rd_order}), we 
cannot obtain the same expression as in Refs. \cite{Bh08,Rom10}. This probably indicates that our
ansatz
Eq.(\ref{Grad}) could be modified. However, an extension of Grad's approach has not been developed
so far.

The main goal of this work is to investigate whether the third-order term
$\propto \pi_{\alpha\beta} \pi^\alpha_\sigma \pi^{\beta\sigma}$ affects local entropy density
production. We expect that this higher-order term reduces the local entropy
density $s$ for typical off-equilibrium initial conditions in heavy-ion collisions, e.g. the color
glass condensate, i.e. it acts in same way as the second-order term introduced earlier by Israel and
Stewart. 

Taking the divergence of the entropy current and using the Gibbs-Duham relation \cite{IS,M04}, we
obtain
\begin{eqnarray}
\partial_\mu s^\mu &=& \frac{1}{T}\pi_{\alpha\beta} \sigma^{ \alpha\beta } - \pi_{\alpha\beta}\pi^{\alpha\beta} \partial_\mu \left( \frac{\beta_2}{2T}u^\mu \right) - \frac{\beta_2}{T}  \pi_{\alpha\beta}\dot{\pi}^{\alpha\beta} \nonumber \,\\
&-& \frac{8}{9}\partial_\mu \left( \frac{\beta_2^2}{T} u^\mu
\right)\pi_{\alpha\beta}\pi^{\alpha}_{\sigma}\pi^{\beta\sigma}-\frac{8}{3}  \frac{\beta_2^2}{T}
\pi_{\alpha\beta}\pi^{\alpha}_{\sigma}\dot{\pi}^{\beta\sigma} \,,
\label{entropy_production_3rd_order}
\end{eqnarray}
where
\begin{equation}
\sigma^{\mu\nu}=\nabla^{\langle\mu}u^{\nu\rangle}=\left(\frac{1}{2}(\Delta_{\alpha}^{\mu}\Delta_{
\beta}^{\nu}
+\Delta_{\alpha}^{\nu}\Delta_{\beta}^{\mu})-\frac{1}{3}\Delta_{\alpha\beta}\Delta^{\mu\nu}
\right)\nabla^{\alpha}u^{\beta}
\end{equation}
and $\Delta_{\alpha\beta}=g_{\alpha\beta}-u_\alpha u_\beta$ with
the metric $g_{\alpha\beta}={\rm diag} (1,-1,-1,-1)$. $\dot{\pi}^{\alpha\beta}$ is the derivative
with respect to $\tau=\sqrt{t^2-z^2}$.

First we demonstrate that the last term in Eq.(\ref{entropy_production_3rd_order}) always has a
positive part, which is not relevant for the derivation of a relaxation equation
for $\pi^{\alpha\beta}$. In the relaxation regime, which is determined by
$\dot{\pi}^{\alpha\beta} \sim -\pi^{\alpha\beta}$, the sign of the product 
$-\pi_{\alpha\beta}\pi^{\alpha}_{\sigma}\dot{\pi}^{\beta\sigma}$ 
is opposite to that of $-\pi_{\alpha\beta} \pi^\alpha_\sigma \pi^{\beta\sigma}$. Thus the
last term in
Eq.(\ref{entropy_production_3rd_order}) is positive. If the system is initially outside the
relaxation regime, the sign of $-\pi_{\alpha\beta}\pi^{\alpha}_{\sigma}\dot{\pi}^{\beta\sigma}$
will be the same as $-\pi_{\alpha\beta} \pi^\alpha_\sigma \pi^{\beta\sigma}$, i.e., 
negative. Then we can separate the last term of
Eq. (\ref{entropy_production_3rd_order}) into two terms:
\begin{eqnarray}
-\frac{8}{3} \frac{\beta_2^2}{T} \pi_{\alpha\beta}\pi^{\alpha}_{\sigma}\dot{\pi}^{\beta\sigma} &=&
-\frac{8}{3} \left(1-\tau_\pi\theta \right) \frac{\beta_2^2}{T}
\pi_{\alpha\beta}\pi^{\alpha}_{\sigma}\dot{\pi}^{\beta\sigma} \nonumber \\
&& -\frac{8}{3}  \tau_\pi\theta  \frac{\beta_2^2}{T}
\pi_{\alpha\beta}\pi^{\alpha}_{\sigma}\dot{\pi}^{\beta\sigma} \,,
\label{dot_pi_term}
\end{eqnarray}
where $\theta=\partial_\mu u^\mu$ denotes the inverse of the expansion scale
and $\tau_\pi$ denotes the intrinsic relaxation time scale on which the
relaxation of the shear pressure toward the Navier-Stokes value sets in. $\tau_\pi\theta$ is thus
the local Knudsen number \cite{HM09,B09}. We now argue that for $\tau_\pi\theta > 1$, the system is
outside the relaxation regime; whereas for $\tau_\pi\theta < 1$ relaxation sets in. Thus the first
term on the right-hand side of Eq. (\ref{dot_pi_term}) is always non-negative, whereas the second
term is first negative ($\tau_\pi\theta > 1$) and then positive ($\tau_\pi\theta < 1$).

The entropy production Eq. (\ref{entropy_production_3rd_order}) can be rewritten as
\begin{eqnarray}
\partial_\mu s^\mu & = & -\frac{8}{3}  \left(1-\tau_\pi\theta \right) \frac{\beta_2^2}{T}
\pi_{\alpha\beta}\pi^{\alpha}_{\sigma}\dot{\pi}^{\beta\sigma}
+\pi_{\alpha\beta} \nonumber \\
& \times& \bigg[ \bigg.  \frac{1}{T}\sigma^{\alpha\beta} - \pi^{\alpha\beta} \partial_\mu \left(
\frac{\beta_2}{2T}u^\mu \right) 
- \frac{\beta_2}{T} \dot{\pi}^{\alpha\beta}  \nonumber \\
&-& \frac{8}{9} \partial_\mu \left( \frac{\beta_2^2}{T} u^\mu \right)\pi^{\langle
\alpha}_{\sigma}\pi^{\sigma\beta\rangle} -\frac{8}{3} \tau_\pi\theta \frac{\beta_2^2}{T}
\pi^{\langle \alpha}_{\sigma}{\dot\pi}^{\sigma\beta\rangle}  \bigg. \bigg] \nonumber\,.\\
\
\label{factorized}
\end{eqnarray}
Note that each term in the brackets has to be 
traceless for further derivations and we have thus applied the projector
operator to the quadratic terms in $\pi^{\alpha\beta}$.
According to the second law of thermodynamics, the entropy production
should be non-negative. This is guaranteed if we impose a linear relation 
between the dissipative flux $\pi^{\alpha\beta}$ and the expression in 
the square brackets, which can be interpreted as a thermodynamic force:
\begin{eqnarray}
\pi^{\alpha\beta}&=& 2\eta T \bigg[ \bigg. \frac{1}{T}\sigma^{\alpha\beta} - 
\pi^{\alpha\beta} \partial_\mu \left( \frac{\beta_2}{2T}u^\mu \right) - \frac{\beta_2}{T}
\dot{\pi}^{\alpha\beta} \nonumber \\
&-& \frac{8}{9} \partial_\mu \left( \frac{\beta_2^2}{T} u^\mu \right)\pi^{\langle
\alpha}_{\sigma}\pi^{\sigma\beta\rangle} 
-\frac{8}{3} \tau_\pi\theta  \frac{\beta_2^2}{T} \pi^{\langle
\alpha}_{\sigma}{\dot\pi}^{\sigma\beta\rangle}\bigg. \bigg]\nonumber\,,\\
\
\label{pimunu3}
\end{eqnarray}
where $\eta$ is the shear viscosity. Note that if the system is initially in the relaxation regime
the last term in Eq.(\ref{pimunu3}) should not be included in the square brackets since it is a
positive contribution to the entropy production [see Eq.(\ref{factorized})]. By separating Eq.
(\ref{factorized}) into two non-negative parts, we have assumed the maximum entropy production,
which corresponds to the fact that interactions among matter constituents tend to drive the
system toward equilibrium as fast as possible. Dividing both sides of Eq.(\ref{pimunu3})
by $2\eta\beta_2$ and using the Israel-Stewart definition of the relaxation time
$\tau_\pi=2\eta\beta_2$ we finally obtain
\begin{eqnarray}
\dot{\pi}^{\alpha\beta} & = & -\frac{\pi^{\alpha\beta}}{\tau_\pi} +
\frac{\sigma^{\alpha\beta}}{\beta_2} 
- \pi^{\alpha\beta} \frac{T}{\beta_2}\partial_\mu \left( \frac{\beta_2}{2T}u^\mu \right) \nonumber
\\
& - & \frac{8}{9} \frac{T}{\beta_2} \partial_\mu \left( \frac{\beta_2^2}{T} u^\mu
\right)\pi^{\langle \alpha}_{\sigma}\pi^{\sigma\beta\rangle} -\frac{8}{3}  \tau_\pi\theta \beta_2
\pi^{\langle \alpha}_{\sigma}{\dot\pi}^{\sigma\beta\rangle} \nonumber .\\
\
\label{3rd_order_pimunu}
\end{eqnarray}
Equation (\ref{3rd_order_pimunu}) presents a novel \textit{third-order} 
evolution equation for the shear tensor. This is the main result of this
work. We note that $\tau_\pi\theta \sim \tau_\pi/\tau$ is of the same order as $\pi^{\alpha
\beta}/T^4$
for large $\tau$. Therefore, the last term in Eq. (\ref{3rd_order_pimunu}) is of fourth order in 
$\pi^{\alpha \beta}/T^4$ for large $\tau$ and should be taken off from a certain time. More
discussion on the higher-order corrections will be given later in this article.

Neglecting the last two terms in Eq. (\ref{3rd_order_pimunu}),
the second-order Israel-Stewart equation is recovered. As mentioned before, it does not contain all
second-order terms found in recent works \cite{Romatschke:2009im,Rom10,Rom09,R09}. 
Considering one-dimensional boost-invariant expansion, the only difference
is the second-order term $\frac{\lambda_1}{\eta^2 \tau_\pi}\pi^{\langle
\alpha}_{\sigma}{\pi}^{\sigma\beta\rangle}$ \cite{Rom09}, which does not appear in our equation.
However,
recently it has been shown in \cite{Mart09} that including this term does not help 
prevent unphysical behavior. We will demonstrate below that this behaviour will be corrected by
the third-order term found by us, which has a form similar to $\frac{\lambda_1}{\eta^2
\tau_\pi}\pi^{\langle\alpha}_{\sigma}{\pi}^{\sigma\beta\rangle}$ but contains an
additional gradient.

In the following we explicitly give the third-order viscous hydrodynamic equations for
a one-dimensional boost-invariant expanding system. In this particular case the heat flow vanishes
as we assumed at the beginning. The four flow velocity is
$u^\mu=\frac{1}{\tau}(t,0,0,z)$ and thus $\theta=\partial_\mu u^\mu=1/\tau$. In the
comoving frame the shear tensor is diagonal, with a positive shear pressure $\pi>0$:
\begin{equation}
\pi_{\mu\nu}={\rm diag} (0,\pi/2,\pi/2,-\pi) .
\label{diag_p}
\end{equation}
The evolution equation for the energy density $e$ and the number density $n$ follow from the
conservation of the energy momentum tensor, $\partial_\nu T^{\nu 0}=0$, and the particle number
conservation, $\partial_\mu N^{\mu}=0$: 
\begin{equation}
\dot e=-\frac{4}{3}\frac{e}{\tau}+\frac{\pi}{\tau} ~~,~~ \dot n=-\frac{n}{\tau}.
\label{e} 
\end{equation}
Using Eqs. (\ref{e}) we obtain
\begin{equation}
\dot\pi=-\frac{\pi}{\tau_\pi}-\frac{4}{3}\frac{\pi}{\tau}+\frac{8}{27}\frac{e}{\tau}-3\frac{\pi^2}{e\tau}-3\frac{\tau_\pi}{\tau}\frac{\pi}{e}\dot\pi
\label{3rd_O_full}
\end{equation}
according to Eq. (\ref{3rd_order_pimunu}) in the comoving frame.

To count the orders in $\pi/e$ we multiply both sides of 
Eq. (\ref{3rd_O_full}) by $\tau_\pi/e$ and obtain
\begin{equation}
\underbrace{\frac{\tau_\pi}{e}\dot\pi}_{\mathcal{O}(2)}=-\underbrace{\frac{\pi}{e}}_{\mathcal{O}(1)}-\underbrace{\frac{4}{3}\frac{\pi}{e}\frac{\tau_\pi}{\tau}}_{\mathcal{O}(2)}+\underbrace{\frac{8}{27}\frac{\tau_\pi}{\tau}}_{\mathcal{O}(1)}-\underbrace{3\frac{\pi^2}{e^2}\frac{\tau_\pi}{\tau}}_{\mathcal{O}(3)}-\underbrace{3\frac{\tau_\pi^2}{\tau} \dot\pi \frac{\pi}{e^2}}_{\mathcal{O}(4)}
\,.
\label{3rd_O_powers}
\end{equation}
We realize that at late times when the relaxation toward equilibrium starts,
$\tau_\pi/\tau$, which is the local Knudsen number, is of the same order 
as $\pi/e$, which can also be interpreted as the inverse 
of the Reynolds number \cite{M04}. Thus, the last term in 
Eq.(\ref{3rd_O_powers}) is of fourth order and has to be omitted for 
consistency. At early times $\tau_\pi/ \tau > 1$ all higher-order
contributions have to be included. For simplicity we neglect the
last term in Eq.(\ref{3rd_O_powers}) for all times and obtain
\begin{equation}
\dot\pi=-\frac{\pi}{\tau_\pi}-\frac{4}{3}\frac{\pi}{\tau}+\frac{8}{27}\frac{e}{\tau}-3\frac{\pi^2}{e\tau} \,.
\label{3rd_O_1D}
\end{equation}
The term $-3\pi^2/(e\tau)$, which gives the difference from the
Israel-Stewart theory, additionally damps the increase of the shear pressure $\pi$. 

Comparing with Eq. (\ref{3rd_O_1D}), one can assume that the higher-order 
terms are of the form $\sim\left(\frac{\pi}{e}\right)^n \frac{e}{\tau}$. 
In practice, we take a heuristic expression
\begin{equation}
\dot\pi=-\frac{\pi}{\tau_\pi}-\frac{4}{3}\frac{\pi}{\tau}+\frac{8}{27}\frac{e}{\tau}-x\frac{\pi^2}{e\tau} \,,
\label{xpi}
\end{equation}
where the number $x$ is supposed to contain approximately all corrections 
of and beyond the third order. A similar approach, an improved hydrodynamic
theory, which includes 
higher-order gradient terms, has been recently proposed \cite{LS09}. 
Equation (\ref{xpi}) should be valid even 
for a free streaming of particles, which is an extreme case with 
an infinite shear viscosity. Thus $\tau_\pi \sim \eta = \infty$.
For a one-dimensional free streaming
with the Bjorken boost invariance the energy density decreases as
$\dot e=-e/\tau$, which leads to $\pi=e/3$ according to Eq. (\ref{e}).
The latter indicates the vanishing of the longitudinal pressure.
Putting these into Eq. (\ref{xpi}) we obtain $x=5/3$.
We see that the damping of $\dot\pi$ becomes weaker when all higher-order
contributions are taken into account. This also indicates that the 
higher-order corrections may have an oscillating behavior.

Using Eqs. (\ref{e}) and (\ref{xpi}) we obtain
\begin{equation}
\frac{\partial}{\partial\tau}\left( \frac{\pi}{e} \right) = 
-\frac{\pi}{e\tau_\pi}+\frac{8}{27}\frac{1}{\tau}
-\frac{\pi^2}{e^2\tau}-x\frac{\pi^2}{e^2\tau} \,.
\end{equation}
For $x=0$, which indicates the transition to the Israel-Stewart theory, 
the derivative of $\pi/e$ at $\pi=e/3$ can be positive for 
sufficiently large $\tau_\pi$ (or $\eta$). This means that $\pi$ can be
larger than $e/3$, which leads to a negative longitudinal pressure\cite{Mart09} and
thus, is unphysical. Note that the negative effective pressure phenomenon has been observed in Ref. 
\cite{Mart09} using the full second-order equations presented in Refs. \cite{R09,Romatschke:2009im}.
In
contrast, for $x=3$ or $x=5/3$ in Eq.(\ref{xpi}) the derivative of 
$\pi/e$ is negative at $\pi=e/3$. We realize that the third-order (and higher-order) corrections
prevent the unphysical behavior
that appears in the second-order theory.

To demonstrate the significance of higher-order corrections to the
Israel-Stewart equation we solve the hydrodynamic equations ({\ref{e}) and (\ref{xpi}) with $x=0$
(IS), $x=3$ (third order), and $x=5/3$ (all orders approximation), respectively. We also compare
these solutions with those calculated from a transport model, the parton cascade BAMPS \cite{XG05},
which has recently been applied to investigate a wide range of phenomena
such as the buildup of the elliptic flow \cite{XG09}, the energy loss
of high-energy gluons \cite{F09}, the extraction of the second-order 
viscosity coefficient \cite{El09}, and the formation and propagation
of shock waves \cite{B09} in ultrarelativistic heavy-ion
collisions.
 
In this work, we implement only elastic collisions with an isotropic 
differential cross section. In this case, the shear viscosity coefficient
can be calculated via \cite{HM09}
\begin{equation}  
\eta=\frac{6}{5}\frac{T}{\sigma_{22}} .
\label{eta}
\end{equation}
Using the equilibrium entropy density $s=4n-n\ln\lambda$ we obtain
the cross section as a function of $\eta/s$:
\begin{equation} 
\sigma_{22}=\frac{6}{5}\left(\frac{\eta}{s}\right)^{-1}\frac{T}{4n-n\ln\lambda} \,.
\label{cs22} 
\end{equation}
We adjust the cross section locally to keep $\eta/s$ constant in the 
BAMPS calculations. In hydrodynamic calculations the shear viscosity is 
given via $\eta=\left(\frac{\eta}{s}\right) (4n-n\ln\lambda)$.
The initial condition of the one-dimensional boost-invariant expansion
is assumed to be a thermal state ($\lambda=1$) with an initial temperature of
$T_0=500$ MeV at the initial time $\tau_0=0.4$ fm/c.

Figure \ref{fig_pressure} presents the time evolutions of the pressure 
isotropy $p_L/p_T=(p-\pi)/(p+\pi/2)$, where $p=e/3$, for various
constant $\eta/s$ values. 
\begin{figure}[t]
\includegraphics[width=8.7cm,angle=0,origin=t]{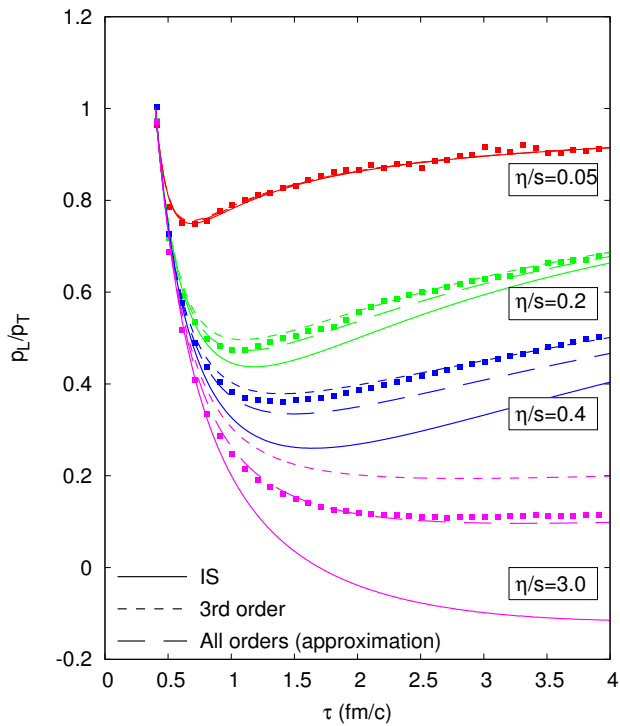}
\caption {(Color online) Time evolution of the pressure isotropy for various $\eta/s$
values. Symbols present the results of BAMPS calculations.
The solid, short dashed and long dashed curves show the solutions
of the Israel-Stewart theory ($x=0$), of the third-order hydrodynamics
($x=3$), and of the heuristic consideration of all-orders contributions
($x=5/3$). $x$ is the parameter in Eq. (\ref{xpi}).}
\label{fig_pressure}
\end{figure}
In all hydrodynamic and transport calculations, the pressure isotropy first decreases from the
equilibrium value $1$ and then turns to increase toward the equilibrium value. The time 
scale of the minimum of the pressure isotropy is proportional to
the relaxation time scale $\tau_\pi=2\eta \beta_2$, which is approximately
proportional to the $\eta/s$ ratio.

Second, the larger the $\eta/s$ ratio, the wider the three
hydrodynamic 
solutions go apart. Whereas for $\eta/s=0.05$ the curves are almost
identical, for $\eta/s=3$ the third-order and approximative all-orders treatments bring
a correction of 300\% and 200\%, respectively, to the result of the
Israel-Stewart theory. [For $\eta/s=1$ (not shown) these corrections are 200\% and 100\%,
respectively.]
The higher-order corrections are essential for
prevention of appearance of unphysical negative pressure, as observed
in the IS solution for $\eta/s=3$.

Third, comparing the results (symbols) from the BAMPS
calculations with 
the solutions (solid lines) of the Israel-Stewart hydrodynamics, we see 
a perfect agreement for $\eta/s=0.05$ and a 10\% difference for $\eta/s=0.2$. 
For $\eta/s=0.4$ the difference is already more pronounced, 30\%; 
and for $\eta/s=3$, it is significant  -- almost 200\%. 
The second-order Israel-Stewart theory breaks down for large $\eta/s$ values.
This is in line with the findings in Refs. \cite{HM09,El09}.

Finally, the hydrodynamic solutions in
the third-order and approximative all-orders 
treatments are much closer to the kinetic transport solutions than the 
second-order ones. The heuristic ansatz for including all orders
corrections leads to remarkably good agreements with the BAMPS results
for $\eta/s$ values in a wide range, although it is hard to extend
this treatment to a general form such as Eq. (\ref{3rd_order_pimunu}).
We see that the main deviation of the third-order results from the
transport ones appears during the early times, at which the local Knudsen
number $\tau_\pi/\tau$ is larger than $1$ and the dissipative effect is 
still strong, so that the third-order treatment
is not sufficient to describe the real hydrodynamic expansion.
This is also the reason why the approximative all-orders treatment provides a somewhat better
description at early times.

If we use a smaller initial time $\tau_0$, which will be the case at the CERN Large Hadron
Collider (LHC), the system would evolve
faster out of equilibrium. Smaller $\eta/s$ values have to be chosen to obtain the same quantitative
behavior as in Fig.\ref{fig_pressure}.

In summary, we have derived a novel third-order 
evolution equation for the shear stress tensor in dissipative hydrodynamics. The higher-order
corrections to the second-order equation from the Israel-Stewart theory have been investigated. We
found that the third-order contribution is essentially needed to prevent unphysical behaviors,
which may occur for large $\eta/s$ ratios. Comparisons between hydrodynamic
and kinetic transport calculations have quantitatively demonstrated the 
significance of higher-order corrections for description of expanding viscous
matters. The solutions of the higher-order hydrodynamic equations showed
good agreement with the kinetic transport results in a wide range of
the $\eta/s$ ratio, although the derived equation does not contain all possible second-order terms
obtained in the literature.

The authors thank D.~H.~Rischke and P.~Huovinen for fruitful discussions and 
comments and for their interest in this work. A.~E. thanks A.~Muronga for discussions and
acknowledges 
the hospitality of UCT, Cape Town, where part of this work was accomplished.
A. ~E. also acknowledges support by the Helmholtz foundation.
This work was supported by the Helmholtz International Center 
for FAIR within the framework of the LOEWE program launched by the State of Hesse.



\end{document}